# Observation of Conservations Laws in Diffusion Limited Aggregation

Mark B. Mineev-Weinstein and Ronnie Mainieri

Center for Nonlinear Studies, MS–B258
Los Alamos National Laboratory
Los Alamos, NM 87545

**Abstract**

We repeat the numerical experiments for diffusion limited aggregation (DLA) and show that there is a potentially infinite set of conserved quantities for the long time asymptotics. We connect these observations with the exact integrability of the continuum limit of the DLA (quasi-static Stefan problem). The conserved quantities of the Stefan problem (harmonic moments) when discretized are our conserved quantities. These numerical experiments show that the exact integrability of the Stefan problem may be continued beyond the formation of cusps in the moving boundary.

Since Witten and Sander discovered the universality of fractal growth in the diffusion limited aggregation (DLA) model [1] much effort has been devoted to developing a theory for it (see, for example, the collections of reprints [2, 3, 4]). The interest stems from the similarities between the DLA model and a variety of physical phenomena such as dielectric breakdown, viscous fingering as in the Hele-Shaw cell, electrodeposition, solidification from supersaturated solutions, and nucleation in the freezing of a liquid [2, 5, 6]. Nevertheless, basic questions remain concerning DLA [7]. We neither know, for example, how to derive the Hausdorff dimension of the DLA-cluster, nor how to calculate correlations.

DLA is a model for particle aggregation in which particles perform a random walk until they collide with an aggregate, upon which they stick to it. As mentioned by Witten and Sander [1], the DLA model is the lattice version of the Stefan problem. Here we assess whether the exact integrability that exists for pattern formation in the continuous two-dimensional Laplacian growth process (the Stefan problem in a quasi-static limit) [8, 9] holds also for DLA. The remarkable result is that despite the stochastic and discrete nature of the DLA model, which excludes the direct use of the integrability results, much of the theory can still be applied.

To understand the motivation for the results presented later, we briefly review the theory of exact integrability for Laplacian growth. The interface $\partial S$ between the aggregate $S$ and a sea of random walkers changes with time. At each instant $t$ one can conformally map the exterior region of the unit circle to the exterior region of $S$ by the map $f(t, \cdot)$. The boundary of the circle $e^{i\phi}$ is mapped to the boundary $\partial S$. The conformal map changes as the interface changes, and its evolution is governed by the nonlinear partial differential equation:

$$\text{Im}(\bar{f}_t f_\phi) = 1 \ . \tag{1}$$

In this equation, $\bar{f}_t$ and $f_\phi$ are partial derivatives, with $f_\phi \neq 0$ outside the unit circle, and the diffusion constant and average concentration of Brownian walkers have been set to one. To our knowledge this equation first appeared in 1945, introduced independently by Galin [10] and Polubarinova-Kochina [11, 12], and rediscovered several times since then.

It turns out that this equation has some remarkable properties. In particular Eq. (1) possess an infinite number of constants of motion [8, 9], which are the harmonic moments of the moving domain. Given that one can further show that for a scalar harmonic field with a source located at infinity the conserved quantities are:

$$m_n = \int_{R^2 \setminus S} \frac{dx \, dy}{(x + iy)^n} \tag{2}$$

The integration is over the exterior domain where the scalar field is harmonic (satisfies the Laplace equation), and $n = 1, 2, \ldots$. These moments have a geometrical interpretation in terms of the Schwarz function describing the shape of the moving boundary [13, 14].

It is believed that all exactly integrable partial differential equations have at least one discrete version which is also exactly integrable [15]. Therefore it is natural to look for an exactly integrable lattice model for which Eq. (1) is the continuum limit. Of course the DLA cannot be such a model because of its non-deterministic nature — the DLA cluster grows as a result of collisions with particles in Brownian motion. Nevertheless, one might expect that while in the initial stages of growth the shape of the DLA cluster is very sensitive to noise, in later stages it is not. We understand it in the sense that quantities that were conserved in the continuous model (the moments $m_n$) would change only very slightly in a lattice DLA-process (at least in asymptotics). So we have decided to repeat the DLA experiments, but rather than calculating a fractal dimension we instead focus on the time-evolution of the discrete version of the moments defined by Eq. (2).

The numerical scheme consists of growing an aggregate on a two-dimensional square lattice $L$ where sites can be occupied or not. The lattice size is $N \times N$ where $N = 601, 901, 2501$ with periodic boundary conditions. The sites that form the aggregate are permanently occupied and do not move. The simulation is started by putting a seed at the center of the lattice, occupying the $(0,0)$ site. Then a random walker is dropped close to the edge of the lattice and allowed to wander the lattice until it collides with the aggregate. At this point the walker is considered part of the aggregate. The distribution of starting points for the walkers is a uniform circle of radius $N/2$ centered at the origin. After the walker becomes part of the aggregate a new walker is released into the lattice and the process is repeated. A typical cluster is shown in Fig. 1a.

In between the release of the walkers the discrete version of the integrals for the moments, Eq. (2), are computed over the sites of the lattice that are not part of the aggregate. The typical behavior of these discrete moments is shown in Fig. 2, where the real part of moments $m_3$, $m_6$, and $m_9$ are plotted as a function of the number of particles in the cluster (time); similar behavior is observed for the imaginary part. The lattice spacing was taken to be 1 in the simulations. Because the process is stochastic, each different simulation yields different values for the moments $m_n$, but the order of magnitude for the moments in Fig. 2 is typical. The mean value of any moment $m_n$, for $n \geq 1$ is zero (as the disk is the average DLA cluster), but the exact distribution of the moments, or even the mean-square value of any moment $m_n$, is not analytically known. As $n$ becomes larger, the variations in the value of a moment becomes smaller, but they are never zero. One can see this by looking at the variations of the value of $m_6$ in Fig. 2d. Results of these calculations show that, after some initial transient time, both real and imaginary parts of the discrete analogs of the moments $m_n$ from Eq. (2) become constant with a high degree of accuracy. This confirms our hypothesis that in a long-term limit discreteness and randomness of the fractal growth are no longer important. Further, the process is governed (at least in asymptotics) by the (potentially) infinite set of constants of motion of the continuous problem.

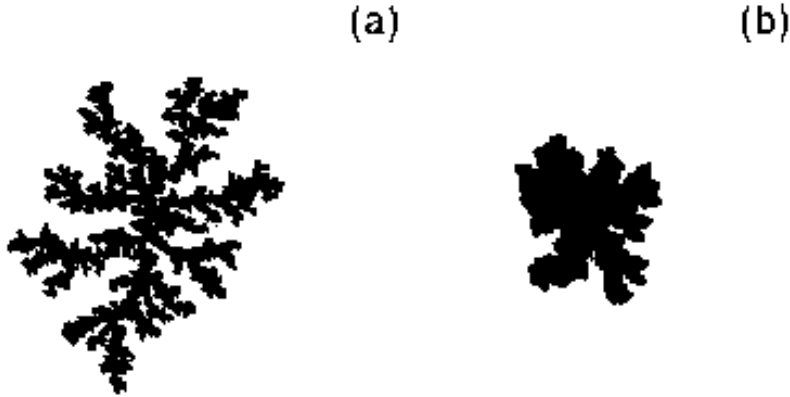

Figure 1: DLA clusters used to compute the moments in Figs. 2 and 3. Cluster (b) is smoother than cluster (a) because it is grown with a simulated surface tension. (This is a low resolution version of the figure. A full resolution version is the file /pub/DLA/dla.ps.Z, which can be ftp'ed from mynah.lanl.gov.)

It is known that the introduction of surface tension in the problem breaks the exact integrability mentioned earlier. In terms of moments it means that they are no longer conserved. To investigate this we introduced in our numerical studies a non-zero surface tension by having some of the particles not stick when they collide with the aggregate cluster. The probability of sticking is a function of the number of occupied neighbors the collision site has. In our simulations we choose 0.08, 0.3, and 1.0 for 1, 2, and 3 occupied neighbors. By giving low probability to colliding with a site with a single neighbor, the DLA cluster tends to be more rounded, simulating the effect of surface tension. This can be seen in Fig. 1, where clusters with and without surface tension are compared. In Fig. 3 a typical moment is computed for a cluster grown with and without surface tension. The moments when there is surface tension are no longer conserved and slowly go to zero. This result is also in agreement with the continuous theory which predicts a decaying magnitude for the moments in the presence of surface tension.

To complete our comparison between the exact integrability results and

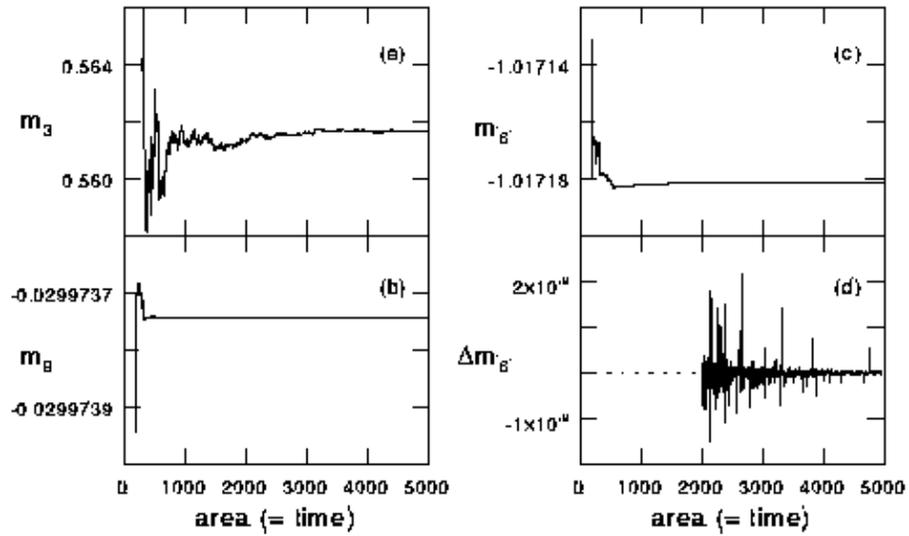

Figure 2: Conserved moments of Eq. 1 as a function of time (area). The first part of each plot is omitted as it varies much more than the asymptotic tail. As the order of the moment increases, its absolute value decreases, as can be seen in (b). Even though the plots appear to be constant, there still is variation, as can be seen in (d) where the variations $\Delta m_6$ of $m_6$ are plotted. (This is a low resolution version of the figure. A full resolution version is the file /pub/DLA/dla.ps.Z, which can be ftp'ed from mynah.lanl.gov.)

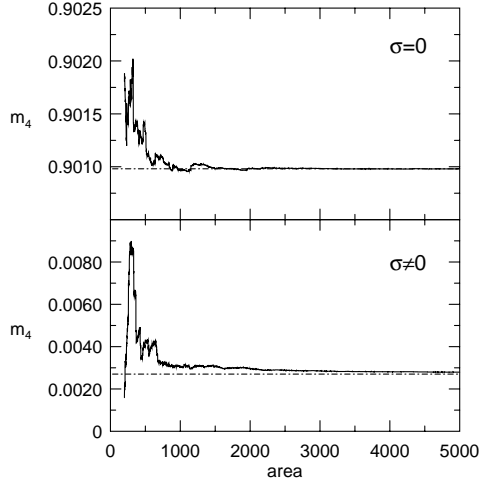

*Figure 3: The moments are trivial if there is surface tension in the problem. Without surface tension, $\sigma = 0$, a moment will tend to a non-zero constant, whereas with surface tension, $\sigma \neq 0$, the same moment will rapidly tend to zero (in the simulations there is always a residual due to the empty (white) regions trapped within the cluster, as in Fig. 1b).*

the DLA model, we will consider the "anharmonic moments"

$$m_{k,l} = \sum_{(x,y) \in L \setminus S} \frac{\cos(k\theta(x,y))}{(x^2 + y^2)^{l/2}} \qquad (3)$$

where the summation takes place over non-occupied sites $L \setminus S$ of the lattice $L$, $k$ and $l$ are positive integers, and $\theta(x,y)$ is the angle between the horizontal and the vector $(x,y)$. We repeated the experiments with the anharmonic moments, Eq. (3). As with the continuous theory, we again find that only the moments $m_{k,l}$ with $k = l$ are conserved.

The variation of anharmonic moments shows that the conservation of $m_n$ is not a numerical accident. As seen in Fig. 2d, the fluctuations of a moment can be quite small when compared to the absolute value of the moment. The possibility then exists that the moment is not really conserved, but varying in a time scale to large to be seen in the numerical experiments. Fig. 4 shows that this is not the case. In all three plots the quantity being summed, $cos(k\theta)/r^l$, falls off with the same power of $r$, the distance from the origin, but only when $k = l$ is the moment conserved. The same behavior is observed for other values of $l$ besides 4. In Fig. 4 the plots have been translated by different amounts but scaled by the same factor. (The units are frac($10^5 m_{k,l}$), where frac gives the fractional part of a number.)

Most initial conditions in Laplacian growth with zero surface tension lead to the formation of cusps in finite time [16]. This limits the applicability of the exact integrability results for the study of the long time limit. However, recently a class of solutions that approximates a broad class having star-like initial condition has been found that does not develop any singularities [17]. Our numerical experiments show that the exact integrability may be continued beyond the formation of cusps also for fractal structures (which are not star-like) such as those in the DLA model.

The main conclusion from this computing experiment is that the DLA possesses a set of conserved quantities in asymptotics in agreement with the continuous quasi-static Stefan problem, and therefore might be integrable (at least partly) in the long-time limit despite its non-deterministic nature. This is a novel view on the processes of fractal growth.

We wish to thank M. Ancona for his useful comment and I. M. Gelfand for his regular interest to this activity and for stimulating discussions.

# References

[1] T. A. Witten and L. M. Sander. Diffusion-limited aggregation, a kinetic critical phenomena. *Physical Review Letters*, 47:1400–1403, 1981.

[2] L. Pietronero and E. Tosatti, editors. *Fractals in physics*. North-Holland, Amsterdam, 1986.

[3] H. E. Stanley and N. Ostrowsky, editors. *On growth and form*. Martinus Nijhoff, Derdrecht, 1985.

[4] R. Pynn and T. Riste, editors. *Time dependent effects in dissolved materials*. Plenum Press, New York, 1987.

[5] P. Pelcé, editor. *Dynamics of curved fronts*. Academic Press, Boston, 1988.

[6] D. Bensimon, L. P. Kadanoff, S. Liang, B. I. Shraiman, and C. Tang. *Reviews of Modern Physics*, 58:977, 1986.

[7] A. Arneodo, F. Argoul, E. Barcy, J. F. Muzzy, and M. Tabard. Golden mean arithmetic in the fractal branching of diffusion-limited aggregates. *Physical Review Letters*, 68:3456–3459, June 1992.

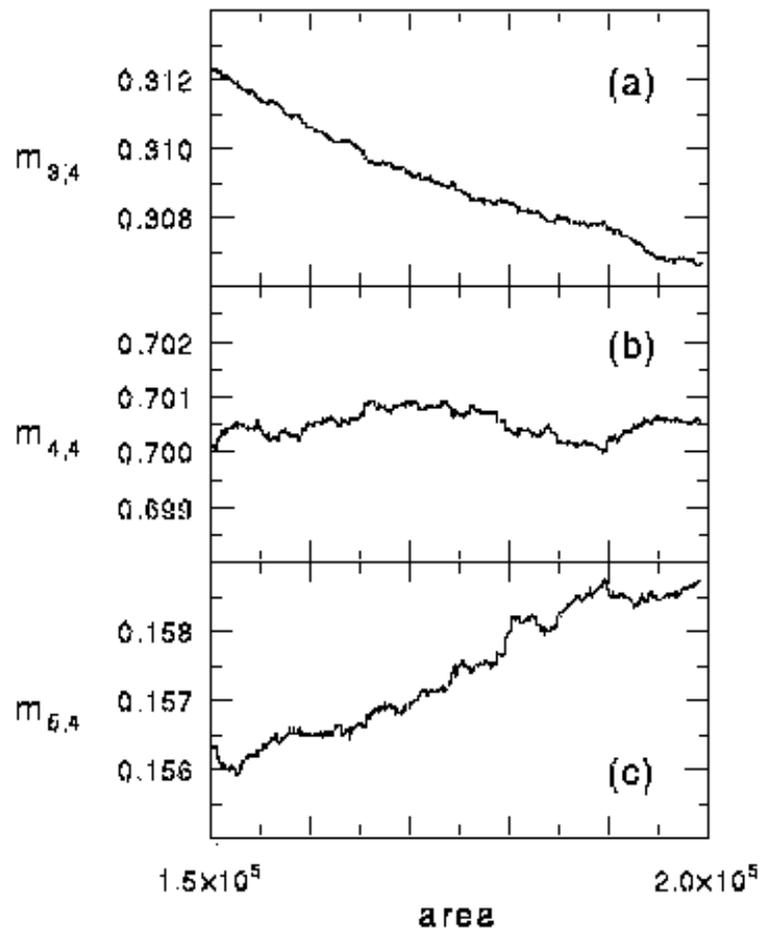

*Figure 4: Anharmonic moments $m_{k,l} = \int \cos(k\theta)/r^l$ after $2 \times 10^5$ time steps. When $k = l$ the moments decay, but if $k \neq l$ a variation in the moments can be seen. The behavior of $m_{k,4}$ is typical. For clarity the initial large variations of each moment are omitted and the vertical units are scaled. (This is a low resolution version of the figure. A full resolution version is the file /pub/DLA/dla.ps.Z, which can be ftp'ed from* `mynah.lanl.gov`.*)*

[8] S. Richardson. Hele-Shaw flow with a free boundary produced by the injection of fluid into a narrow channel. *Journal of Fluid Mechanics*, 56:609–618, 1972.

[9] M. B. Mineev. A finite polynomial solution of two-dimensional interface dynamics. *Physica D*, 43:288–292, 1990.

[10] L. A. Galin. *Dokl. Akad. Nauk. S.S.S.R.*, 47:246–249, 1945.

[11] P. Ya. Polubarinova-Kochina. *Dokl. Akad. Nauk. SSSR*, 47:254–257, 1945.

[12] P. Ya. Polubarinova-Kochina. *Prikl. Matem. Mech.*, 9:79–90, 1945.

[13] M. Mineev-Weinstein. Schwarz function and Laplacian growth. Unpublished, 1992.

[14] S. D. Howison. Complex variable method in the Hele-Shaw problem. preprint, Oxford Mathematical Institute, 1992.

[15] Mark J. Ablowitz. Private communication.

[16] S. D. Howison. Cusp development in the Hele-Shaw flow with a free surface. *SIAM Journal of Applied Mathematics*, 46:20–26, 1986.

[17] Mark B. Mineev-Weinstein and Silvina Ponce Dawson. A new class of nonsingular exact solutions for Laplacian pattern formation. Technical Report LAUR - 93-1127, Los Alamos National Laboratory, 1993.